\documentclass[acmsmall,manuscript,nonacm]{acmart}
\usepackage{hyperref}
\usepackage{hyperxmp}
\usepackage{pifont}
\usepackage{multirow} 
\AtBeginDocument{%
  \providecommand\BibTeX{{%
    \normalfont B\kern-0.5em{\scshape i\kern-0.25em b}\kern-0.8em\TeX}}}

\setcopyright{acmcopyright}
\copyrightyear{2024}
\acmYear{2024}
\acmDOI{XXXXXXX.XXXXXXX}

\acmBooktitle{CSCW '25: The 28th ACM SIGCHI Conference on Computer-Supported Cooperative Work \& Social Computing}
\begin{document}

%\title{QualiGPT: GPT as an Easy-to-use Tool for Qualitative Coding}
\title{When Qualitative Research Meets Large Language Model: Exploring the Potential of QualiGPT as a Tool for Qualitative Coding}
\author{He Zhang}
\email{hpz5211@psu.edu}
\orcid{0000-0002-8169-1653}
\affiliation{%
  \institution{College of Information Sciences and Technology, Penn State University}
  \city{University Park}
  \state{Pennsylvania}
  \country{USA}
  \postcode{16802}
}
\author{Chuhao Wu}
\email{cjw6297@psu.edu}
\orcid{0000-0002-3862-560X}
\affiliation{%
  \institution{College of Information Sciences and Technology, Penn State University}
  \city{University Park}
  \state{Pennsylvania}
  \country{USA}
  \postcode{16802}
}
\author{Jingyi Xie}
\email{jzx5099@psu.edu}
\orcid{0000-0002-2753-2360}
\affiliation{%
  \institution{College of Information Sciences and Technology, Penn State University}
  \city{University Park}
  \state{Pennsylvania}
  \country{USA}
  \postcode{16802}
}
\author{Fiona Rubino}
\email{far5185@psu.edu}
\orcid{0009-0002-9414-4824}
\affiliation{%
  \institution{College of Engineering, Penn State University}
  \city{University Park}
  \state{Pennsylvania}
  \country{USA}
  \postcode{16802}
}
\author{Sydney Graver}
\email{sjg6347@psu.edu}
\orcid{0009-0007-7901-397X}
\affiliation{%
  \institution{College of Engineering, Penn State University}
  \city{University Park}
  \state{Pennsylvania}
  \country{USA}
  \postcode{16802}
}
\author{ChanMin Kim}
\email{cmk604@psu.edu}
\orcid{0000-0001-9383-8846}
\affiliation{%
  \institution{College of Education, Penn State University}
  \city{University Park}
  \state{Pennsylvania}
  \country{USA}
  \postcode{16802}
}
\author{John M. Carroll}
\orcid{0000-0001-5189-337X}
\email{jmc56@psu.edu}
\affiliation{%
  \institution{College of Information Sciences and Technology, Penn State University}
  \city{University Park}
  \state{Pennsylvania}
  \country{USA}
  \postcode{16802}
}
\author{Jie Cai}
\authornote{Corresponding author.}
\orcid{0000-0002-0582-555X}
\email{jpc6982@psu.edu}
\affiliation{%
  \institution{College of Information Sciences and Technology, Penn State University}
  \city{University Park}
  \state{Pennsylvania}
  \country{USA}
  \postcode{16802}
}

\renewcommand{\shortauthors}{Zhang et al.}

\begin{abstract}

Qualitative research, renowned for its in-depth exploration of complex phenomena, often involves time-intensive analysis, particularly during the coding stage. Existing software for qualitative evaluation frequently lacks automatic coding capabilities, user-friendliness, and cost-effectiveness. The advent of Large Language Models (LLMs) like GPT-3 and its successors marks a transformative era for enhancing qualitative analysis. This paper introduces QualiGPT, a tool developed to address the challenges associated with using ChatGPT for qualitative analysis. Through a comparative analysis of traditional manual coding and QualiGPT's performance on both simulated and real datasets, incorporating both inductive and deductive coding approaches, we demonstrate that QualiGPT significantly improves the qualitative analysis process. Our findings show that QualiGPT enhances efficiency, transparency, and accessibility in qualitative coding. The tool's performance was evaluated using inter-rater reliability (IRR) measures, with results indicating substantial agreement between human coders and QualiGPT in various coding scenarios. In addition, we also discuss the implications of integrating AI into qualitative research workflows and outline future directions for enhancing human-AI collaboration in this field. %QualiGPT represents a significant step towards more efficient, accessible, and rigorous qualitative research in the era of artificial intelligence.
\end{abstract}


% gpt, tool, qualitative

\begin{teaserfigure}
  %\centering
  \includegraphics[width=1\linewidth]{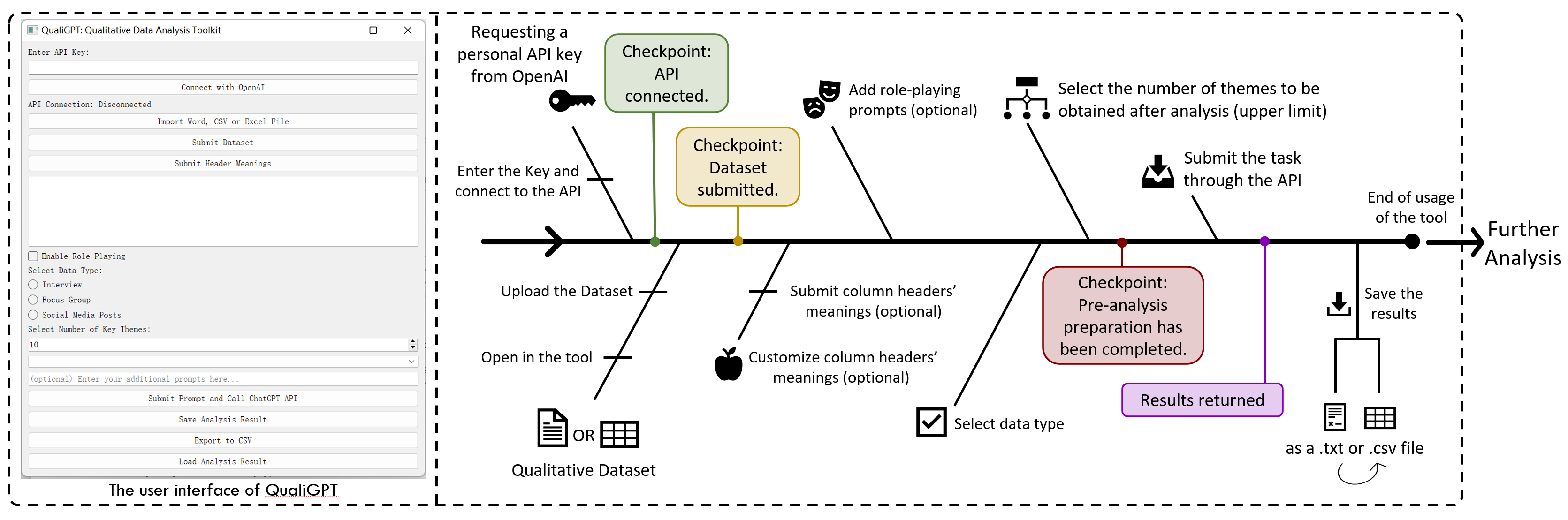}
  \caption{\textbf{Overview of the qualitative analysis toolkit, QualiGPT. The user interface of QualiGPT is displayed on the left. On the right side, the usage flow and design logic of QualiGPT are presented.}}
  \label{fig.teaserfigure}
\end{teaserfigure}


%\received{1 July 2024}

\maketitle
%%%% [TRACK] Blending Interaction: Engineering Interactive Systems & Tools
%%%% [TRACK] Computational Interaction

\section{Introduction}
%motivation
Qualitative research provides a unique perspective into individuals' comprehension, attitudes, and insights regarding technology, phenomena, and specific topics. Over time, an increasing number of researchers have acknowledged the significance of qualitative methodologies across diverse fields. In the CSCW (Computer-Supported Cooperative Work) field, qualitative research methods are particularly important because they provide deep insights into how people collaborate with the support of technology. CSCW researchers often use qualitative methods such as interviews, observations, and content analysis to understand complex socio-technical systems and work practices. However, as the scale and complexity of data increase, CSCW researchers face the challenge of efficiently processing and analyzing large volumes of qualitative data, after all, analyzing qualitative data can be labor-intensive~\cite{10.1145/3449168}, especially with extensive and complex datasets. Moreover, the task of coding qualitative data not only demands significant effort but also poses challenges related to understanding context and ensuring consistency. Coding, arguably the most crucial task in qualitative analysis, is both a beloved and challenging aspect for analysts. Continuously optimizing methods for processing qualitative data remains a common goal among these professionals. As the production of qualitative data continues to surge, there is an escalating demand for innovative techniques to streamline and enhance the thematic analysis process~\cite{bazeley2013qualitative}.

To address these challenges, researchers have ventured into the development and utilization of qualitative analysis software. These tools employ computer-assisted collaborative efforts to simplify data management and enhance efficiency~\cite{elliott2021exploring}. While such software has indeed streamlined the coding process and improved the quality of coding to some extent, existing platforms like Nvivo\footnote{https://lumivero.com/products/nvivo/} and atlas.ti\footnote{https://atlasti.com/} still have limitations in terms of performance and operational complexity, failing to fully meet the needs of researchers~\cite{hansson2010multiple,bergin2011nvivo}.

In addition to the high subscription costs, learning to use these software tools for qualitative data analysis is not straightforward. Early-career researchers or analysts often find themselves investing a significant amount of time in understanding how to accomplish their target tasks within these environments and navigating the multifaceted UI interfaces~\cite{10.1145/3492844}. However, many of these features are designed to cater to specific needs. In other words, not all functionalities within the software are utilized frequently by analysts, leading to increased learning overheads. As described by Ragavan et al.~\cite{10.1145/3490099.3511161}, analysts not only have to be concerned about their primary tasks at hand but also bear the additional learning costs associated with the tools (software) they choose. Therefore, the development of a more user-friendly tool to reduce the workload of analysts in their primary workflows becomes especially crucial.
%-
Starting from 2022, with the emergence of GPT-3, researchers began to widely recognize the immense potential of Large Language Models (LLMs) in various domains. The subsequent releases of GPT-3.5, GPT-4 and GPT-4o, were perceived by many as heralding a comprehensive technological revolution. The advent of large-scale language models seemed to offer a new avenue of possibilities. It was during this time that we were inspired to ponder whether it might be feasible to leverage LLMs to assist in qualitative analysis, aiming to enhance both efficiency and performance. To achieve this objective, we approached it from a practical standpoint, selecting one of the most popular LLM applications developed by OpenAI, ChatGPT and its API, as our research platform to bolster the universality of our research contributions. We drew inspiration from the recent works~\cite{zhang2023redefining,10.1145/3617362,10.1145/3584931.3607500} on enhancing qualitative analysis using ChatGPT, emphasizing the importance of prompts and by extending some of the future work they had highlighted.

In summary, this study first categorizes and summarizes the typical issues encountered when using ChatGPT, identifying four major categories that encompass eight common types of erroneous ChatGPT responses. Concurrently, we compiled concerns from previous studies wherein analysts expressed reservations about employing ChatGPT for qualitative analysis tasks, as well as the challenges ChatGPT faces in such contexts. With these issues and challenges in mind, we introduced QualiGPT: a user-friendly integrated tool built on API and prompt design, specifically tailored for thematic analysis of qualitative data. The tool's performance was evaluated using inter-rater reliability (IRR) measures~\cite{mchugh2012interrater}, with results indicating substantial agreement between human coders and QualiGPT in various coding scenarios.

We deployed QualiGPT~\cite{zhang2023qualigptgpteasytousetool} on both simulated and real datasets and compared its performance to manual coding. The results show that this tool effectively addresses the challenges inherent in the traditional qualitative data coding process. It streamlines the qualitative analysis workflow, reduces costs associated with processing qualitative data, and alleviates concerns regarding transparency and credibility in using ChatGPT for qualitative analysis. Additionally, due to its integrated design and API implementation, QualiGPT offers marked improvements in usability, user-friendliness, privacy protection, and performance over the web version of ChatGPT. When compared to conventional software, QualiGPT provides a more insightful user interface, significantly lowering the learning and usage costs for researchers.

\section{Related Work}

Qualitative research is an important way to understand the human world. This method is based on experience and subjectivity, maintaining an open stance towards the meanings of the research subject through practical interactivity~\cite{doi:10.1080/09518390600975701}, and it is widely used by researchers across various disciplines. In the process of qualitative research, data analysis and criteria for rigor form an important component~\cite{6581909a-2d2f-3026-8aca-a6cd9c090b74}. Researchers need to remain sensitive to the data and provide discussion and insights through continuous interpretive and critical thinking. Coding, as a method for analyzing qualitative data, is crucial for searching concepts, ideas, themes, and categories within the data to help researchers organize and interpret the data~\cite{given-sage-2024}. As~\citet{saldana2015coding} stated, no one can have the ultimate authority over the ``best'' method for qualitative coding due to the inherently high flexibility of qualitative analysis. Therefore, in this section, we will quickly review some of the processes and methods of qualitative coding and summarize some common coding processes. In addition, we will review a series of challenges that exist in qualitative coding. Based on our research objectives, we will examine both manual coding and coding conducted with technological assistance.

\subsection{Manual Qualitative Coding}
Coding is the most crucial part of qualitative research, and the process is often labor-intensive~\cite{10.1145/3449168,elliott2018thinking,zhang2009qualitative}. Although coding enhances the understanding of data, and researchers can gain a series of new insights from the coding process, human coders or annotators still typically need to spend a considerable amount of time on the manual coding process and face a range of challenges~\cite{zhang2023redefining}. To further address or reduce the negative impact of the challenges encountered during the coding process, we first briefly review the inductive and deductive coding methods and their associated challenges.

\subsubsection{Inductive Coding}
Inductive coding is one of the commonly used methods in qualitative analysis~\cite{thomas2003general}. This approach is based on the data itself, extracting and inferring themes through the analysis of the data~\cite{forman2007qualitative}. In this section, we will briefly review the processes and roles of human coders and AI in inductive coding.

In qualitative inductive coding, human coders play a vital role. By reviewing qualitative data, human coders are tasked with identifying key concepts, themes, or patterns that emerge from the data, without relying on any pre-established codebook~\cite{fereday2006demonstrating}. This approach shares similarities with the process of open coding, as both do not rely on pre-existing theories or assumptions and maintain flexibility throughout the coding process to adjust coding strategies based on the data~\cite{naeem2023step}. This suggests that coders often require a certain level of coding experience or need to work under the guidance of experts~\cite{wiltshire2021realist}.

Inductive coding can be particularly challenging for novice coders, as qualitative data is often more complex and messier compared to quantitative data~\cite{kalman2019requires}. Human coders must ensure coding consistency and identify implicit concepts and themes in the inductive coding process, which demands strong reflective abilities and skills, such as memo-writing~\cite{stuckey2015second}.

Moreover, since inductive coding is typically an iterative process, coders are required to manage the data and the entire project~\cite{noble2015issues}. When collaborating with others in the coding and analysis stages, coders also face challenges such as maintaining an open mindset, avoiding over-interpretation, and possessing high management skills.

\subsubsection{Deductive Coding}
Deductive coding is another primary method in qualitative coding~\cite{fereday2006demonstrating}. In deductive coding, human annotators interpret the data using a pre-established coding framework, applying codes from a codebook~\cite{macqueen1998codebook} to the data~\cite{given-sage-2024}. Once a certain number of codes have been generated, deductive coding can be applied~\cite{kennedy2018deduction}. This approach requires human annotators to carefully review the qualitative data and the codebook, assigning codes to segments of the qualitative data that correspond to the predefined categories in the codebook~\cite{goodell2016practical}. This process demands a lower level of experience in exploring implicit patterns in the data compared to inductive coding. However, it necessitates a higher understanding of each coding category and its connotations~\cite{azungah2018qualitative,graneheim2017methodological}.

Consistency is equally crucial in deductive coding, and maintaining coding consistency becomes a significant challenge~\cite{o2020intercoder}. Annotators typically need to ensure the consistency of their coding decisions through meetings, discussions, agreements, and technical methods, promoting a shared understanding of the codes and the content being coded~\cite{10.1145/3173574.3173733,hemmler2022beyond,chowdhury2015coding}. Multiple rounds of discussions among coders are particularly important for resolving issues encountered during the deductive coding process.

When human annotators encounter data segments that do not fit neatly into the existing coding framework, they must decide how to handle these ``exceptions,'' either by creating new codes or adjusting the existing ones~\cite{pearse2019illustration}. Multiple rounds of discussions among coders are particularly important for resolving issues encountered during the deductive coding process, which has been evidenced in several previous studies~\cite{10.1145/3290607.3312879,lyu2024preliminary,10.1145/3613904.3642787}.

\subsection{Technology-Assisted Qualitative Coding}

Although researchers in qualitative research can choose different coding methods based on various data types, research purposes, and research processes, these coding methods often present a series of challenges for human coders, such as manual labor costs~\cite{10.1145/3449168,elliott2018thinking,zhang2009qualitative}, consistency~\cite{o2020intercoder} and reliability issues~\cite{golafshani2003understanding}, coding management~\cite{belotto2018data}, coding experience~\cite{holton2007coding}, and the learning curve for novices~\cite{peredaryenko2013calibrating}. To address these challenges, qualitative researchers have been continuously striving to improve the performance of qualitative coding and tackle the challenges encountered during the qualitative coding process through technological means.

With the increasing amount of qualitative data being generated, Computer-Assisted Qualitative Data Analysis (CAQDA) has been playing a critical role in qualitative research  \cite{pilkington1996Use,weitzman1995computer,chandra2019ComputerAssisted}. CAQDA software nowadays offer a wide range of functionalities such as processing data from multiple media channels (text, picture, audio, and video), visualizing the analysis results through automatic plotting of data, and quickly generating predefined and customized reports \cite{phillips2018quick}. While the capabilities vary greatly among applications, more advanced functionalities often comes at the cost of a high subscription fee \cite{renaissancerachel202315} that potentially deter researchers away. As a result, some free and open-source alternatives have been developed to support the growing need of qualitative research, such as Taguette \cite{rampin2021Taguette} and RQDA \cite{chandra2019overview}, although their functionalities tend to be more basic than commercial products. Another problem with CAQDA software is their user experience and learnability. \citet{paulus2013Digital} find that initial encounters can be intimidating for novices, yet with proper guidance, researchers can effectively integrate these tools. Still, studies on the interface design of CAQDA are rather limited and a comparison of both commercial and open source applications is necessary for designing better tools. 

The combination of AI and qualitative research has begun to redefine how researchers approach qualitative data and analysis~\cite{haque2022think,10.1145/3581754.3584136}. Technologies, especially AI algorithms, provide potential for improved efficiency in analyzing large datasets, a task that traditionally requires substantial time and resources when conducted by human analysts. In fact, in earlier years, researchers have been using computers or technologies to assist in qualitative studies~\cite{coffey1996qualitative,weitzman1995computer,thunberg2022pioneering}. 

AI can be used to gather and organize qualitative data from various sources, like social media platforms, online forums, and digital archives. This not only saves time and resources but can also uncover a wider range of data points that might be overlooked in manual collection~\cite{feng2023investigating}. Also, AI-powered transcription services can transcribe audio and video data into text format quickly and accurately. Typically, transcription and encoding in qualitative research present the biggest challenges for researchers, often consuming a lot of time. However, a good assistant tool allows researchers to focus more on analysis rather than on data preparation~\cite{10.1145/3173574.3173922}. AI models can provide an initial analysis of textual data by summarizing content, identifying key themes, sentiments, or trends, and even insightful advice and generating questions that can help guide further research~\cite{cui2023survey,ma2020machine,khurana2023natural,shaik2022review,10.1145/3411764.3445591}. By comparing AI findings with human analysis, researchers can increase the validity and reliability of their findings~\cite{10.1145/3544548.3581352}. With AI's ability to process data rapidly, researchers can conduct real-time analysis during data collection, helping them adjust their research approach as needed based on preliminary findings~\cite{panda2019artificial}. 

\subsubsection{LLMs and Prompt Engineering}
% what is prompt engineering
% chatgpt is one example
The advent of automated qualitative analysis techniques has enabled qualitative researchers to analyze volumes of data that would be difficult to analyze manually~\cite{welsh2002dealing}, and the rise of LLMs may further enhance the efficiency of analysis.

Prompt engineering is the deliberate design and optimization of instructions, or ``prompts'', aimed at enhancing the performance and accuracy of LLMs when generating outputs~\cite{reynolds2021prompt,zamfirescu2023johnny}. This strategy is crucial as the type and specificity of prompts provided to LLMs can significantly shape their responses.

ChatGPT\footnote{https://openai.com/chatgpt} by OpenAI\footnote{https://openai.com/}, developed within the Generative Pretrained Transformer (GPT) framework, underscores the importance of prompt engineering~\cite{fiannaca2023Programming}. It is capable of understanding and generating human-like text (natural language), offering an interactive experience similar to that of human interactions. Renowned for its expertise in diverse language tasks, such as producing human-like text, content generation, sentence completion, and in-depth essay or report writing~\cite{liebrenz2023generating,alkaissi2023artificial,bishop2023computer,macdonald2023can}. However, ChatGPT is not immune to errors. It may yield outputs that seem nonsensical or incorrect, particularly when faced with unclear or ambiguous prompts~\cite{shen2023chatgpt,hassani2023role}. Hence, applying prompt engineering to enhance the capabilities of LLMs is a crucial method.

The value of prompt engineering gains further emphasis from studies revealing improved outcomes when LLMs like ChatGPT receive meticulously crafted prompts. Techniques such as few-shot learning~\cite{zhao2021Calibrate}, chain-of-thought methods~\cite{wei2022ChainofThought}, and role-playing scenarios~\cite{gao2023Prompt} have demonstrated considerable efficacy. However, the performance of ChatGPT, even when paired with refined prompt engineering, can differ based on the domain in question. Mastery in domain-specific knowledge is pivotal for honing the model's efficacy~\cite{tian2023Opportunities, wang2023Brief}. Thus, practitioners are encouraged to weigh the specific application context carefully during prompt engineering~\cite{heston2023Prompt}.
For areas like qualitative analysis, employing an iterative methodology—consistently adapting and evaluating diverse prompt engineering strategies—may be instrumental in harnessing the full potential of ChatGPT~\cite{10.1145/3617362,zhang2023redefining,10.1145/3584931.3607500}.

Despite the impressive capabilities of AI, machine learning, and LLMs, the complex nature of qualitative analysis presents unique challenges that these technologies are still learning to navigate~\cite{10.1145/3544548.3580688}.

%%%%%%%%%%
\section{Overall Motivation and Design Considerations of QualiGPT}
By leveraging techniques proposed by researchers for using ChatGPT in qualitative task analysis, we tested and refined these techniques on the web version of ChatGPT. We integrated the solutions into our toolkit, serving as resources and prior knowledge for the development of QualiGPT. Specifically, the design considerations encompass two main parts: the first pertains to the common concerns of qualitative analysts about applying ChatGPT to qualitative analysis tasks, as introduced in Section~\ref{challengesofchatgpt}. The second pertains to some of the current shortcomings of the web version of ChatGPT, as discussed in Section~\ref{initialperformance}.

Our development of QualiGPT was informed by testing and refining techniques proposed by researchers for utilizing ChatGPT in qualitative task analysis. We conducted these tests on the web version of ChatGPT and integrated the resulting solutions into our toolkit, serving as resources and prior knowledge for QualiGPT's development. The design considerations encompass two main areas: first, addressing common concerns of qualitative analysts regarding the application of ChatGPT to qualitative analysis tasks, as introduced in Section~\ref{challengesofchatgpt}; and second, addressing some of the current limitations of the web version of ChatGPT, as discussed in Section~\ref{initialperformance}.

\subsection{Common Challenges and Concerns of Qualitative Analysis and the Use of ChatGPT in the Qualitative Analysis Process}\label{challengesofchatgpt}

In previous researches~\cite{10.1145/3584931.3607500,zhang2023redefining}, scholars have identified several significant challenges associated with incorporating LLMs into qualitative data analysis. We revisited these challenges and further explored how to address them by designing an integrated tool. We revisited these challenges and further contemplated how to address them through the design of an integrated tool. In summary, our design aims to tackle the following challenges: (1) \textbf{Lack of Transparency}: ChatGPT's ``black box'' nature makes it difficult for researchers to understand how it processes data. Better prompt design can improve interpretability and transparency, (2) \textbf{Consistency and Context Understanding Issues}: Varied responses and difficulty in maintaining context in multi-turn dialogues are key challenges. Streamlined and precise prompts can enhance consistency, (3) \textbf{Difficulty in Prompt Design}: Creating effective prompts is time-consuming and lacks a standardized approach. QualiGPT simplifies this process with pre-designed prompts, (4) \textbf{Challenges in Understanding ChatGPT's Responses}: Using ChatGPT for qualitative analysis doesn't always save time compared to traditional methods, especially if its outputs are lengthy and disorganized. This contradicts the goal of improving efficiency. To address this, we can design prompts to standardize ChatGPT's outputs for better readability. Additionally, implementing strategies to prioritize reading sequences can help users quickly identify key concepts, enhancing overall efficiency, and (5) \textbf{Data Privacy and Security}: In the digital era, data privacy is a major concern, especially when using large language models like ChatGPT for research. The risk of sensitive data leakage is significant, as shown by historical data breaches~\cite{6246173}. While efforts like GDPR~\cite{voigt2017eu} and encryption techniques~\cite{salavi2019survey} are made to protect data privacy, individual users also need to be aware of their online security practices. According to OpenAI's policies, user data in ChatGPT can be used to improve the model, while API service data isn't used for model training. This suggests that the API service offers better data security~\cite{mather2009cloud}.

\subsection{ChatGPT Initial Performance Test}\label{initialperformance}
During the design process, we analyzed 1,000 data entries from a public Discord channel to assess ChatGPT's ability to provide accurate and standardized content. We also summarized typical errors, examples, and potential solutions encountered during the testing process. These primarily include: (1) network errors, (2) incorrect handling of data, (3) Violation of policy, and (4) Out of limits. It's worth noting that these errors are not exclusive to using ChatGPT for qualitative analysis tasks and have a certain degree of universality across various applications.

\section{Design of QualiGPT} %method - design the tool
\begin{figure}[htbp]
  \centering
  \includegraphics[width=1\linewidth]{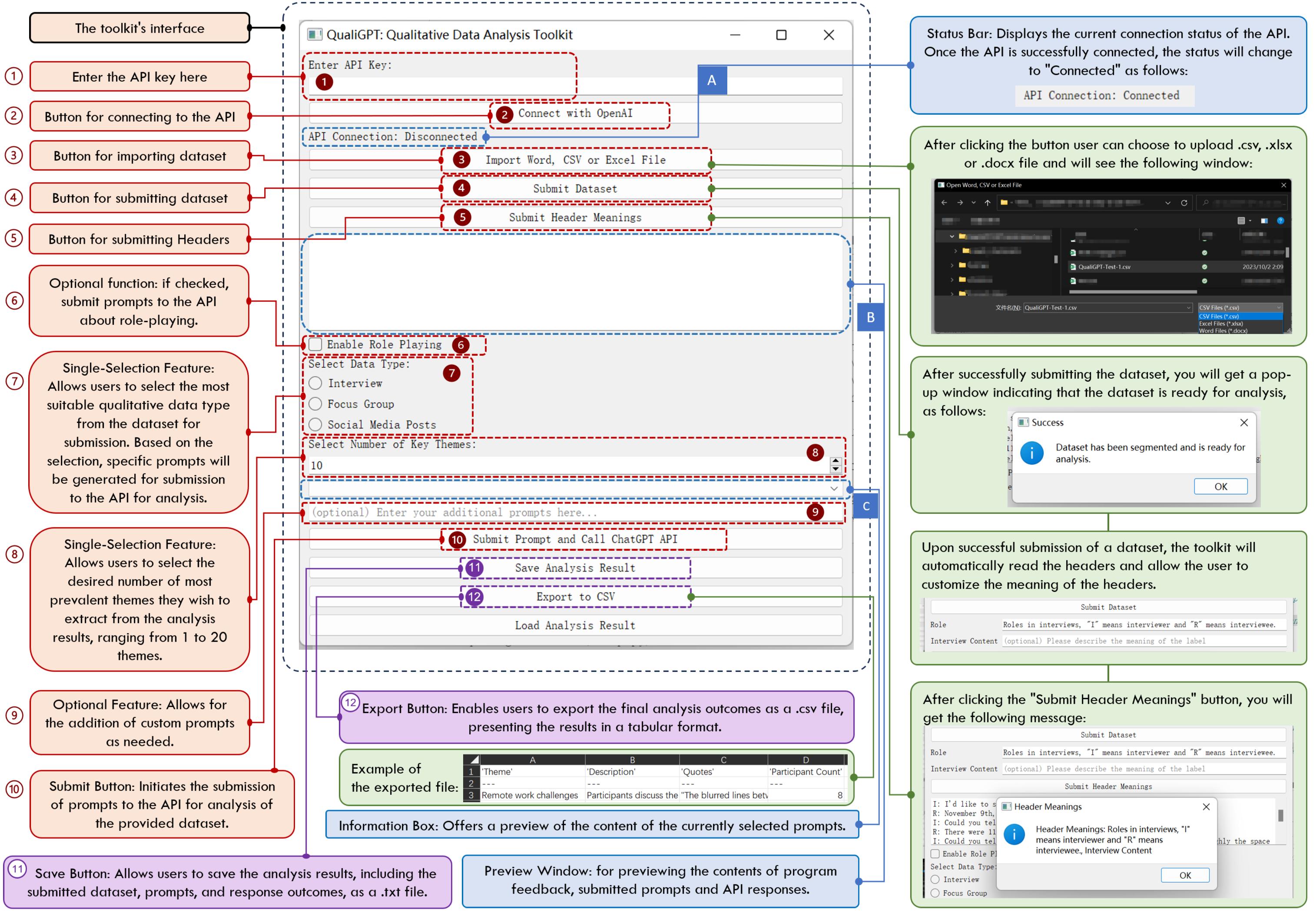}
  \caption{\textbf{User Manual for QualiGPT (A Qualitative Analysis Toolkit) - Interactive Features. QualiGPT offers a total of 13 interactive features that users can select, click, or input text into. The functionalities enclosed by the red boxes are related to invoking the API, while the interactive features shown in the purple boxes do not involve API calls.}}
  \label{fig.QualiGPT}
\end{figure}

To further benefit qualitative researchers, address the challenges presented in Section 3, and overcome the limitations of using ChatGPT on the web interface, we introduce QualiGPT. It's a meticulously crafted, integrated qualitative analysis tool based on prompt engineering and API. This tool features a user-friendly visual interface and is designed to be easily used even by those with no programming experience. Fig.~\ref{fig.teaserfigure} presents the user interface and usage flow of QualiGPT. Fig.~\ref{fig.QualiGPT} is the user manual for QualiGPT. In the following sections of this chapter, we will delve into the functionalities, advantages, and design considerations of QualiGPT.

\subsection{Design Practice}
QualiGPT is designed with a user-centric approach, addressing the challenges researchers face in qualitative analysis and simplifying interactions for novices with LLMs. It integrates OpenAI's API and data processing libraries, providing a simple, user-friendly interface that requires minimal technical expertise. Users only need to provide their API key to access the functionalities of different versions of GPT. Prioritizing user privacy, QualiGPT allows individuals to control their API connections, ensuring secure data exchange and automatic data deletion post-session. Understanding the diversity of qualitative data, QualiGPT is flexible in processing various text formats. It accommodates user-specific parameters like role labels, ensuring analyses align with the dataset's nature. Its dynamic prompt generation, based on research and literature, combines user input with qualitative research principles. QualiGPT's goal is to deliver insights that are comprehensible, shareable, and actionable. The tool presents results in a clear, transferable format, including themes, descriptions, quotes, and participant counts, capturing the essence of qualitative research. With export options, QualiGPT further emphasizes practicality and convenience. The functionalities are detailed in the subsequent sections.

\subsection{Components and Architecture}
\subsubsection{API Connection}
QualiGPT integrates the OpenAI API and Python libraries for advanced text data processing. Users provide their API key for GPT access via a user-friendly interface. This setup bypasses the token input limit of traditional ChatGPT by segmenting data, allowing for more comprehensive analysis. The API's flexibility also enhances data privacy and security, giving users more control over their data.
%QualiGPT operates by harnessing the OpenAI API to access the capabilities of GPT. Users need to setup their own API accounts with OpenAI and provide the key to QualiGPT (\ding{172}, \ding{173}). This setup bypasses the 4096-token input limit of traditional ChatGPT by segmenting data, allowing for more comprehensive analysis. The API's flexibility also enhances data privacy and security, giving users more control over their data.

\subsubsection{User Input and Data Formatting}
QualiGPT supports various textual data formats like Word, .txt, .csv, and .xlsx. Users submit datasets in these formats, receiving a system prompt confirming successful data submission. Optimal processing requires labeled data to differentiate participants in conversations. Users can define roles (e.g., interviewer, interviewee) and provide conversation overviews, which are integrated into a series of prompts to guide GPT in tailored data analysis.
%QualiGPT processes textual data in formats like Word, .txt, .csv, and .xlsx (\ding{174}). Users submit datasets with labels to differentiate participants and provide header meanings for accurate processing (\ding{175}, \ding{176}). Inputs like role labels and conversation descriptions are integrated into prompts for nuanced GPT analysis.

\subsubsection{Prompt Generation}
Prompt generation in QualiGPT is rooted in previous researches~\cite{10.1145/3584931.3607500,zhang2023redefining}, focusing on four components: Task Background, Task Description, Processing Method, and Expected Output. The prompts are customizable, with options like role-playing for expert analysis and selection of data types (e.g., Interviews, Focus Groups). Users can control the depth of analysis by specifying the number of key themes and adding additional instructions. The system offers ``fixed'', ``dynamic'', and ``user-choice-based'' prompts, allowing diverse customization based on user needs. The process culminates in submitting these prompts to GPT for analysis.
%Prompt generation in QualiGPT, guided by Zhang et al.'s research~\cite{zhang2023redefining}, involves four key components: task background, task description, processing method, and expected results, all based on user input. Users can customize prompts with options like role-playing (\ding{177}) and selecting data types (\ding{178}). They can also set the number of themes for extraction (\ding{179}) and add additional instructions (\ding{180}). The submission of these configured prompts for GPT analysis occurs with a click (\ding{181}).
%
%QualiGPT uses three types of prompts: fixed, dynamic, and user-choice-based. Fixed prompts are preset in the code, dynamic prompts are tailored by users, and user-choice-based prompts offer pre-set options for customization. The interplay between these prompts is depicted in Fig.~\ref{fig.prompt-design}.

\subsubsection{Analysis Results}
The prompts generated by QualiGPT guide GPT to execute a qualitative analysis on the submitted dataset, ensuring a rigorous and insightful analytic process. In addition, they guide GPT to present its results in a streamlined, coherent format, tailored for user-friendly interpretation and data exports. Specifically, QualiGPT organizes the results into a tabular format that encapsulates thematic findings. Each table features four columns for `Themes' which represent the overarching patterns or topics within the data. Following that is the ``Description'', illustrating the nuances and depths of these themes. To give a clearer context, the table also includes ``Quotes'' linked to each theme, showcasing direct excerpts from the dataset that support or explain the theme. Moreover, a ``Participant Count'' associated with each theme is presented, offering a quantitative insight into the theme's prevalence or significance. QualiGPT provides users with a practical tool to export these findings in a csv file format. This facilitates further analysis, sharing, or integration with other tools or databases. Additionally, for those keen on preserving the entire analytical journey, from the raw dataset, the constructed prompts, to the derived findings, QualiGPT offers an option to encapsulate all these elements into a singular txt file, ensuring comprehensive documentation and easy recall.

\section{Analysis and Verification}
To demonstrate the performance of QualiGPT, we applied it to both simulated data and real datasets. By comparing the topics returned by QualiGPT with manually coded results, we showcased the powerful potential of QualiGPT in qualitative data coding tasks. In the LLM-assisted coding process, we designed prompts based on the recommendations from past researches~\cite{zhang2023redefining,10.1145/3584931.3607500}. The prompts included descriptions of the task background, objectives, methods, outputs, and inputs, incorporating role-playing and acknowledgment expressions. Modular features, such as role-playing and acknowledgments, are included in the predefined prompts of QualiGPT, while more specific and customized information, like task background, can be input through text boxes.

\subsection{Case Study One - Exploring Themes and The Advantages of Integrated Tools}
In our case study one, we used ChatGPT to create a simulated focus group dataset on ``transitioning to remote work''. The dataset comprises 9,309 words, averaging about 27 words per feedback. It includes 6 medium-length responses (average 112 words) and 2 long responses (average 391 words). This dataset provides diverse insights into the experiences of transitioning to remote work, highlighting both benefits like flexibility and challenges like work-life balance and technical issues. Researchers agreed that the dataset offers a broad thematic variety, reflecting a range of personal experiences and strategies related to remote work from individuals of various backgrounds and job roles.

\subsubsection{Results and Evaluation}
We submitted the data to both ChatGPT (web version) and QualiGPT. In QualiGPT, we selected the data type as ``focus group'' and enabled the ``role-playing'' feature. We also chose to obtain 20 potential topics. The final response results from QualiGPT and ChatGPT (web version) were similar. However, when using the web version of ChatGPT, we encountered several troubling issues that were resolved in QualiGPT:
\begin{itemize}
    \item[1.] Due to the data volume exceeding the token limit for a single submission, we had to manually split the dataset and input it into ChatGPT using the copy-paste method.
    \item[2.] On the web version, we had to manually input prompts multiple times for debugging.
    \item[3.] We had to manually organize the output results, such as transferring the results to a spreadsheet. 
\end{itemize}
This made the work time and complexity on the web version of ChatGPT much greater than using QualiGPT.
To highlight the efficiency advantage of QualiGPT, we repeated the same analysis process in QualiGPT three times and timed each run. From entering the API (starting checkpoint) to saving to a .csv file (ending checkpoint), the results showed that the average time to complete the analysis process in QualiGPT was 96.5 seconds. We provide the simulated dataset used for testing in the supplementary materials, and we welcome researchers to use this dataset for a quick test on QualiGPT to experience the efficiency improvement compared to the web version of ChatGPT or manual coding.

\subsection{Case Study Two - Inductive Coing by using LLM}
%%% 不要引用你的paper
In Case Study 2, we used a real social media dataset collected in the past study and conducted inductive coding on 200 entries using both manual coding and LLM-assisted coding methods.

For manual inductive coding, we performed topic modeling on the raw dataset, which resulted in 8 distinct topics. We selected data from two of these topics and randomly extracted 100 entries from each. These entries were then independently subjected to inductive coding by two research assistants. Each entry was labeled with tags consisting of 2-5 words. Each entry in the dataset is a message from a user in that channel. The dataset does not contain any identifiable information. In the first round of manual coding, coding 200 entries took several hours of work, and the entire manual coding process lasted close to a week including discussions and negotiations on coding. 

For inductive coding by using LLM, we removed the manually coded labels from the data and submitted it to QualiGPT for analysis, and we designed prompts that allow QualiGPT to explore the data that was processed in the manual inductive coding section and generate corresponding codes without prior knowledge. Specifically, we interact with QualiGPT through prompts. First, we employ role-playing to activate QualiGPT's capabilities, such as ``\textit{You are now an excellent qualitative data analyst and qualitative research expert.}'' Then, we inform it about the task it needs to assist with and provide the task background, for example, ``\textit{You need to perform inductive coding on a dataset that was obtained from a public Discord server named `TwitchDev'. This server is run by non-staff volunteers such as moderators and administrators. TPDs (Twitch Platform Developers) will get a developer role in this Discord by the administrators of this Discord server if they prove that their program developments are building for Twitch users or using the Twitch-provided tools. The community has thousands of TPDs to share their experiences. It has more than 2,000 active members daily, including Twitch official staff, TPDs, broadcasters, and viewers.}'' We also inform it about the format of the input data, such as ``\textit{Each row in the dataset represents a single data entry,}'' and specify the output format, ``\textit{Please help me determine a possible code for each data entry and return the results in a tabular format, with the first column being the data index and the second column being the code.}'' Since GPT can only input a limited number of prompts (tokens) in each interaction, we additionally employ prompts with acknowledgment to enhance the power of context in memory between each iterations and inputting new data, such as ``\textit{Great job! Please continue analyzing the following data: [New Data].}'' If GPT provides an incorrect format or is not processing the task correctly, we will regenerate it.

\subsubsection{Results and Evaluation}
After the coding was completed, we aimed to verify the IRR between human and LLM in the inductive coding task. We re-evaluated the consistency between the coding results produced by the two research assistants, after discussion and negotiation, and those produced by GPT-4. This re-evaluation was necessary because the coding was done without a prior codebook, leading to potential variations in vocabulary usage. Specifically, we marked data entries as ``1'' for consistency if their meanings were identical or similar, such as ``twitch merch'' (human-coded label) and ``Twitch merchandise'' (LLM-coded label). Conversely, entries were marked as ``0'' for inconsistency if their meanings differed or if they represented different levels, such as ``reputation of developer'' (human-coded label) and ``Twitch user behavior and reputation'' (LLM-coded label). After this annotation process, we calculated Cohen's Kappa, which resulted in a value of 0.57.

Overall, the LLM exceeded expectations in completing the inductive coding task. Achieving high consistency among different coders is particularly challenging without extensive prior knowledge~\cite{fereday2006demonstrating}, as each individual may have their own interpretation of the data~\cite{saldana2015coding}.

\subsection{Case Study Three - Deductive Coding by using LLM}

\subsubsection{Codebook development}
Before conducting deductive coding, we first needed to prepare a codebook. This codebook was developed based on the inductive coding results from Case 2. Specifically, the two research assistants compared the labels for each data entry, either agreeing on one of the two labels or combining them to develop more suitable labels. The negotiated labels were then compiled into the initial codebook.

The senior researcher and the two research assistants reviewed this initial codebook, which contained 171 labels. They removed labels that were not relevant to the theme of social support. This refinement process resulted in a final codebook comprising 54 labels, where label 0 indicated that the message's topic was irrelevant, and label 53 indicated that the topic was relevant but not specified by the other labels. These labels became the codes that were going to be used to categorize in deductive coding process.

\subsubsection{Deductive Coding Process}
%The research assistants then individually coded another 200 random data entries by using the codebook that had been created. Coders then had a discussion meeting to reach an agreement on how each comment should be coded.

Subsequently, the research assistants used the created codebook to independently code another 200 randomly selected data entries. Simultaneously, we changed the prompt description of task to deductive coding and asked GPT-4 to code the same 200 data entries using the codes from the codebook. To minimize the impact of randomness in LLMs, we had GPT-4 perform three rounds of deductive coding on the 200 data entries, with each round being conducted independently. We also tested the deductive coding on the latest models (GPT-4o and Claude 3.5). The coding results and corresponding data indices from each round were also stored in a table for subsequent comparative analysis.

\subsubsection{Results and Evaluation}

Upon completing the coding process, we calculated the IRR between human coders and LLMs for the deductive coding task. The results indicated that the kappa value between human coders was approximately 0.73, reflecting substantial agreement. The Fleiss' Kappa value between human coders and GPT-4 ranged from 0.44 to 0.50, with an average of approximately 0.46, signifying moderate agreement. Among the three independent GPT-4 coding results, the Fleiss' Kappa value was approximately 0.87, demonstrating almost perfect agreement. Additionally, we tested newer models, GPT-4o and Claude 3.5, which showed Fleiss' Kappa values with human coders ranging from 0.38 to 0.42, indicating moderate agreement. Additionally, we restarted the LLM process and randomly selected 50 data entries, along with their codes, which had been agreed upon by the researchers. These entries were used as prior knowledge and provided to the LLM (GPT-4) through prompting. The LLM then performed deductive coding on an additional 150 data entries. The results showed that the Fleiss' Kappa value between the LLM and the two human coders was approximately 0.46. However, because the kappa value between the two research assistants decreased to 0.64, the consistency between the LLM and the researchers slightly increased. Detailed results are presented in Tabl~\ref{tab:kappa}.

% \usepackage{multirow}
% \usepackage{graphicx}
\begin{table}[htbp]
\centering
\caption{IRR for Inductive and Deductive Coding Across Various Coders}
\label{tab:kappa}
\resizebox{\columnwidth}{!}{%
\begin{tabular}{ccccc}
\hline
\textbf{Index} & \textbf{Type of Coding}               & \textbf{Coders}      & \textbf{Number of Coders} & \textbf{Kappa Value$^{\alpha}$} \\ \hline
1              & Inductive Coding                      & [Human coders], GPT-4      & 2                         & 0.57                 \\ \hline
2              & \multirow{5}{*}{Deductive Coding}     & RA1, RA2             & 2                         & 0.73                 \\
3              &                                       & RA1, RA2, GPT-4      & 3                         & 0.44 - 0.50          \\
4              &                                       & GPT-4 (s)              & 3                         & 0.87                 \\
5              &                                       & RA1, RA2, GPT-4o     & 3                         & 0.38                 \\
6              &                                       & RA1, RA2, Claude 3.5 & 3                         & 0.42                 \\ \hline
7              & Deductive Coding with prior knowledge & RA1, RA2, GPT-4 & 3                         & 0.42                 \\ \hline
\multicolumn{5}{l}{\begin{tabular}[c]{@{}l@{}}$^{\alpha}$ Round to two decimal places\end{tabular}}
\end{tabular}%
}
\end{table}

Overall, in the deductive coding phase, although the LLM did not surpass human researchers in consistency, the results still demonstrated significant potential. Notably, when we used different LLMs for coding, the IRR among the LLMs was very high. This indicates that LLMs perform more consistently than human coders when using a codebook for deductive coding.

\section{Discussion}
In recent times, the advent and evolution of LLMs such as GPT-3.5 Turbo and GPT-4 have opened up new avenues for automating tasks that were traditionally labor-intensive. One such task is the coding of qualitative data to derive thematic insights. Our tool, QualiGPT, leverages the capabilities of LLMs through prompt design and API calls to automate this coding process, offering a list of potential themes. This integrated tool significantly reduces the overhead associated with manual coding, addressing challenges encountered in traditional qualitative analysis and when using ChatGPT.

Specifically, QualiGPT employs prompts that have been validated in prior research~\cite{zhang2023redefining}, offering researchers an efficient means of categorizing themes in qualitative data. The prompts are highly structured, mitigating risks associated with using GPT for analysis, such as inconsistencies and lack of transparency. Compared to traditional qualitative analysis methods or software, the computational prowess of LLMs ensures that QualiGPT outperforms conventional software's auto-coding features in terms of performance. Furthermore, its coding speed far surpasses manual coding while maintaining a quality comparable to expert groups. This tool has the potential to revolutionize the paradigm of qualitative analysis in the future. In this section, we delve into the contributions and prospects of this tool, especially in terms of collaboration.

\subsection{Reflections on LLM-Assisted Qualitative Coding}
Our study provides valuable insights into the potential of LLMs in qualitative research, particularly in the realm of inductive and deductive coding. The findings from our analysis and verification process reveal several key points worthy of discussion.

\subsubsection{Human Consistency vs. LLM Potential}
Our results demonstrate a high level of consistency between human coders, with Cohen's Kappa values reaching 0.73 in deductive coding tasks. This underscores the current gold standard in qualitative analysis. However, it's crucial to note the promising performance of LLMs, particularly GPT-4, in these same tasks. While not yet matching human-level consistency, GPT-4 achieved moderate agreement with human coders, with Fleiss' Kappa values ranging from 0.44 to 0.50. This performance is particularly noteworthy given the complexity of qualitative coding tasks. Typically, as the number of independent coders increases, consistency tends to decrease significantly in the beginning. Therefore, the current stage of IRR should be considered acceptable. Additionally, in multi-coder tasks, low consistency is not a severe issue because consensus can be reached through multiple rounds of group discussions and negotiations. During these discussions, different coders usually need to articulate their viewpoints, which helps foster mutual understanding and enhance the depth of analysis. In future work, if researchers use LLMs for qualitative coding, we recommend more interaction with the LLM. For example, researchers could ask the LLM to provide detailed considerations and explanations for the coded content and share their thought processes during coding with the LLM. Currently, such discussions with LLMs are largely constrained by the interaction method (through text prompts). However, in the future, this could be achieved through multimodal interactions~\cite{xie2024emergingpracticeslargemultimodal,Tai_2024_CVPR} and enhanced context memory and understanding in subsequent LLM versions. Additionally, as LLM continues to evolve rapidly, we anticipate significant improvements in their coding capabilities. The consistently high agreement between different LLM iterations (Fleiss' Kappa of 0.87) suggests that as these models become more sophisticated, they may approach or even surpass human-level consistency in qualitative coding tasks. In the future, we may need to pay more attention to the depth of analysis provided by LLMs in coding.

\subsubsection{Practical Application of LLMs in Qualitative Analysis}
Unlike previous studies that have explored high-level applications of LLMs in qualitative research (such as generating broad themes or categories), our work delves into the practical application of LLMs for specific coding tasks. We demonstrate the capability of LLMs to perform both inductive and deductive coding, core processes in qualitative analysis. In inductive coding, where no prior codebook exists, GPT-4 showed promising results with a Cohen's Kappa of 0.57 when compared to human coders. This suggests that LLMs can effectively identify emergent themes and patterns in qualitative data, a task traditionally requiring significant human expertise and intuition. For deductive coding, where a predefined codebook is used, GPT-4's performance was also encouraging. As we mentioned above, the moderate agreement with human coders indicates that LLMs can effectively apply predetermined codes to new data, a crucial skill in many qualitative research projects. Additionally, during the coding process, researchers need to repeatedly review the data and use methods such as note-taking to reinforce their familiarity with the data context, codebook, and specific coding content. This practice enhances consistency in the coding process but also poses a challenge to the researchers' expertise. This challenge is particularly pronounced when coding large datasets over extended periods (e.g., several weeks) and when multiple rounds of group coding are required, as well as when coding segmented data (e.g., interviews with Q\&A or social media data). Variations in the data and new information emerging during the coding process may affect the consistency of the coders' perspectives. Although the effectiveness of LLMs in coding all data types is not yet clear, our work demonstrates the capability of LLMs to code more structured data. This capability allows researchers to process large amounts of data more quickly and has the potential to identify content that human coders may overlook.

\subsection{QualiGPT as a Tool: Leveraging QualiGPT to Augment Efficiency in Qualitative Analysis}
A primary concern among researchers regarding GPT-generated content stems from a lack of confidence in its accuracy~\cite{poldrack2023aiassisted}. There have been instances where GPT has been found to fabricate content, generating spurious information. Such behavior is unequivocally unacceptable in scientific research. However, when used as an auxiliary tool, these concerns can be significantly alleviated. In other words, when used as a tool, QualiGPT merely offers perspectives on the data, while the mechanism for human review remains intact. Under this modality, researchers can utilize QualiGPT for rapid coding. Specifically, they can select themes of interest from the generated responses and, aided by the justifications provided by QualiGPT (explanations and references to the original data), manually verify these themes. In this scenario, the researcher or user retains control over the accuracy of the results, with the final decision-making power remaining human-centric.

\subsection{QualiGPT as a Collaborative Researcher}
Qualitative analysis often carries a degree of subjectivity, which is typically viewed as an advantage~\cite{garcia1997qualitative,ratner2002subjectivity}, allowing for unique insights to be gleaned from the data~\cite{mohajan2018qualitative}. Concurrently, this subjectivity can lead to varied interpretations of the same qualitative data by different researchers. In traditional analysis workflows, discussions between co-researchers to reconcile coding results and reach a consensus are indispensable. Building on this procedural concept, we pondered the possibility of incorporating QualiGPT as an independent co-researcher in studies.

Under this new paradigm, both human researchers and LLM would analyze the qualitative data independently. Once the analyses are completed, the results from both the human researchers and LLM would be collated for a collective discussion, aiming to achieve consensus among all parties. Indeed, LLM appears to possess the potential to facilitate such a collaborative model, as it can generate several high-quality themes, providing genuine content references from the original text for each theme. Additionally, the advanced version of the LLM supports voice interaction, which introduces new opportunities for qualitative analysis in collaboration with AI. This enhancement makes the LLM function more like a research participant~\cite{dillion2023can}. In this context, LLM should not merely be perceived as a tool assisting researchers but rather as an independent contributor, offering insights into the data and actively participating in discussions.

\subsection{Challenges and Considerations}
Despite the promising results, it's crucial to approach the use of LLMs in qualitative analysis with caution. Several challenges and considerations emerge from our findings: (1) \textbf{Consistency Across LLM Versions:} While we observed high consistency between different iterations of GPT-4, the performance varied when testing other models like GPT-4o and Claude 3.5. This highlights the importance of model selection and the need for researchers to validate results across different LLM versions. However, although our results show that newer versions of LLMs have lowered the IRR with human coders, this does not necessarily mean we should avoid using the updated models. This is because IRR is not the sole criterion for assessing the quality of analysis in qualitative research, and discrepancies in IRR can be addressed through discussion and iteration. Additionally, various test results~\cite{xia2024leaderboardrankingcoding,openai-gpt4o-2024may,anthropic_claude35} indicate that newer LLMs generally exhibit better reasoning abilities, speed, accuracy, and a larger knowledge base, supporting more diverse content (different languages, multimodal data, etc.). The issue of lower IRR with newer models might be due to LLMs producing more varied interpretations of the data, which can be further explained through discussions about inconsistent coding with the LLM. Additionally, the prompts we used may not be perfect, potentially not fully activating the LLM's capabilities or affecting its judgment. To address this, we have retained a customizable prompt window in QualiGPT to support more tailored needs. (2) \textbf{Codebook Development:} Our results showed improved performance when using a predefined codebook for deductive coding. This underscores the continued importance of human expertise in developing and refining coding frameworks, even as LLMs take on more coding tasks. Also, when we attempted to provide more contextual information (prior knowledge) for GPT to learn from. However, while the results showed a slight improvement, it was not significant. This suggests that such learning may need to be built on more complex prompt-guided interactions, such as enabling the LLM to gain a more detailed understanding of the codebook development process. However, at present, this process may be complicated due to the challenge of (3) \textbf{Human-AI Communication.} Unlike human coding teams who can meet to discuss inconsistencies and reach consensus, current LLM implementations lack this interactive capability. Developing tools and methodologies for effective human-AI communication in the coding process remains a challenge. On one hand, the lack of more human-like interaction methods required for more complex interactions is a challenge. In human communication, people can exchange and understand large amounts of information in a short time through speaking, listening, reading, and writing. However, in current LLM interactions, typing (writing) prompts via a keyboard remains the primary mode of interaction, which significantly hinders the efficiency of information transfer. Hence, in the future, enabling LLMs to participate in discussions and negotiations through voice interaction could greatly assist in establishing prior knowledge for LLMs. On the other hand, understanding context is a crucial aspect of effective human communication. Currently, LLMs' understanding of context does not yet reach a human-like level~\cite{zhang2024futurelearninglargelanguage}, which can result in LLMs "forgetting" previous content during the communication process, thereby affecting their task performance. However, trends in LLM iterations show that newer models are progressively improving their memory of context. This is one of the key reasons why we remain open to using the latest versions of LLMs. (4) \textbf{Ethical Considerations:} Although QualiGPT was developed based on previous research findings and conceptualizations, and we believe it has achieved a high degree of usability and user-friendliness in addressing certain practical issues, it is not without flaws. This is particularly important to note given that LLM-assisted collaborative coding is still in its early stages. Our use of APIs is also aimed at strengthening ethical and policy considerations, which aligns with the consistent goals of CSCW. (5) \textbf{Validation and Oversight:} While LLMs show promise as coding assistants, human validation and oversight remain crucial. Researchers should view LLMs as tools to augment, not replace, human analysis in qualitative research.  An innovative direction is to have LLMs critique their own content. This concept involves multiple LLMs process analyzing and debating perspectives, with a human-in-the-loop for final review and decision-making.

\section{Conclusion}
The realm of qualitative research, while invaluable for its depth and nuance, has long grappled with the challenges of data analysis, particularly during the coding phase. Traditional qualitative analysis software, despite their merits, often fall short in addressing the complexities, costs, and performance demands of modern research. This study has illuminated a promising avenue for the future of qualitative analysis through the integration of LLMs, specifically GPT and its API, into the research workflow.

Our introduction of QualiGPT represents a significant stride forward in addressing the longstanding challenges in qualitative data analysis. By identifying and addressing the common issues associated with ChatGPT, we have not only enhanced the efficiency of the coding process but also bolstered the credibility and transparency of using LLMs in qualitative research. The comparative analysis between QualiGPT and manual coding underscores the tool's potential in streamlining the workflow, reducing processing costs, and ensuring a more transparent and credible analysis process.

Furthermore, the design considerations of QualiGPT, with its emphasis on usability and user-friendliness, mark a departure from the often cumbersome interfaces of traditional qualitative software. By offering a more intuitive interface, QualiGPT significantly diminishes the learning and usage overheads, making it an attractive option for both seasoned researchers and those in the early stages of their careers.

In light of our findings, it is evident that the integration of LLMs like ChatGPT into qualitative research holds substantial promise. As technology continues to evolve, it is imperative for the academic community to remain adaptive and open to such innovations, ensuring that research methodologies are not only rigorous but also efficient and user-centric. With tools like QualiGPT, we are one step closer to realizing this vision, ushering in a new era of qualitative research that marries depth with efficiency. Future work should continue to refine and expand upon these tools, ensuring they remain relevant and effective in the ever-evolving landscape of qualitative research.

\begin{acks}

\end{acks}
\appendix
\section{More Figures}

\section{Online Resources}
%We will open source QualiGPT upon acceptance of the paper.
QualiGPT is open-sourced on Github (\url{https://github.com/KindOPSTAR/QualiGPT}).%KindOPSTAR/QualiGPT}.

\bibliographystyle{ACM-Reference-Format}
\bibliography{sample-base,notes}


\end{document}